# Evidence for spin swapping from modulation of transverse resistance in magnetic heterostructures with Rashba interface

Heeman Kim[1], Shutaro Karube[1], Juan Borge[2], Junyeon Kim[3*],

Kouta Kondou[3], YoshiChika Otani[1,3*]

[1] *Institute for Solid State Physics, University of Tokyo, Kashiwa, Chiba 277-8581, Japan*

[2] *Nano-Bio Spectroscopy group, Departamento de Fisica de Materiales, Universidad del Pais Vasco UPV/EHU, E-200018 San Sebastian, Spain.*

[3] *Center for Emergent Matter Science, RIKEN, Wako, Saitama 351-0198, Japan*

**We investigate the transverse response under the out-of-plane magnetic field for magnetic heterostructures with Cu/$Bi_2O_3$ or Ag/$Bi_2O_3$ Rashba interfaces. We detect opposite contributions on the transverse resistance by the Cu/$Bi_2O_3$ and the Ag/$Bi_2O_3$ interfaces, which interestingly coincide well with the opposite signs of the spin/charge interconversion from the two interfaces. We suppose the opposite influences on the transverse resistance feature spin swapping occurrence of the converted spin current. The transverse spin flow emerges due to the spin swapping in both Cu and Ag layer, but the direction of the spin flow is opposed dependent on the spin direction of the converted spin current.**

In solid systems, the spin/charge interconversion takes place as a result of the spin Hall effect (SHE) in three-dimensional bulks[1], and/or the Edelstein effect (EE) at two-dimensional interfaces[2, 3]. The spin/charge interconversion offers a novel playground for innovative applications mediated by spins[4]. It provides an alternative mechanism for high-speed and low power assumption magnetic memories[5-9] and efficient gigahertz nano-oscillators[10]. Moreover, it opens a novel way for the generation of the electricity from various physical quantities such as light[11, 12], heat[13], and mechanical vibration[14-17].

Interestingly, the spin/charge interconversion also significantly alters the magneto-transport properties. The longitudinal resistance is modulated dependent on the magnetization orientation due to the spin/current interconversion. Depending on the origin of the spin/charge conversion or the spin transport mechanism, the modulation of the longitudinal resistance is called spin Hall magnetoresistance (SMR)[18-20], Edelstein magnetoresistance (EdMR)[21, 22], and unidirectional magnetoresistance (UMR)[23-25]. Likewise, the SMR and the EdMR provide a modulation of the transverse resistance identical to symmetry of the anomalous or the planar Hall effects[19, 20, 26]. Alternatively, a recent theoretical study proposes that scattering of the spin current dependent on ferromagnet's magnetization orientation induces a transverse response identical to the anomalous Hall effect (AHE)[27]. Additionally, the interfacial scattering originated from a heavy element composed capping layer significantly modulates the AHE[28].

Here we present the modulation of the anomalous Hall type transverse resistance caused by spin current. In this study, we utilize Cu/$Bi_2O_3$ and Ag/$Bi_2O_3$ Rashba interfaces for the spin current generation. Interestingly, the two Rashba interfaces bring the opposite modulation of transverse resistance correlating with their opposite spin chirality. We consider that the spin swapping effect of the spin current[29] makes a transverse spin current, and consecutively it makes spin chirality dependent transverse resistance modulation.

To investigate the influence of spin/charge interconversion on the transverse resistance ($R_{xy}$), we prepared $Ni_{80}Fe_{20}$ (Py) (10 nm)/Cu (0-20 nm)/$Bi_2O_3$ (20 nm) (Sample 1) and Py (10 nm)/Ag (0-20 nm)/$Bi_2O_3$ (20 nm) (Sample 2) magnetic heterostructures. Here the Cu/$Bi_2O_3$ and the Ag/$Bi_2O_3$ interfaces are Rashba interfaces with the opposite spin chirality (opposite sign of Rashba parameter $\alpha_R$). A previous spin transport study verified the Cu/$Bi_2O_3$ and Ag/$Bi_2O_3$ interfaces have opposite signs of the spin/charge interconversion[30]. Photoemission spectroscopy studies also display the opposite Rashba splitting at the Cu/Bi and the Ag/Bi interfaces[31, 32]. As a reference, we also prepared Py (10 nm)/Cu (0-20 nm) (Reference 1) and Py (10 nm)/Ag (0-20 nm) (Reference 2). All films were patterned to Hall bar devices with 4.0 μm width and 120 μm length (Fig. 1(a)). And all structures are deposited using the electron-beam evaporator, and fabricated using the photolithography and lift-off process.

The measurement was carried out by applying the external magnetic-field along to the out-of-plane (z-axis in Fig. 1(a)). During the measurement, we swept the magnetic field from –6 T to +6 T and applied a 100 μA probe current. All the measurements were performed in the physical property measurement system. Figures 1(b) and (c) display typical $R_{xy}$ for Samples 1 and 2 as a function of the field. Note that the displayed results do not include the contribution of the ordinary Hall effect in the non-magnetic metal (NM) layer (Cu or Ag). We obtain the difference of $R_{xy}$ ($\Delta R_{xy}$) under positive ($R_{xy}$(+H)) and negative ($R_{xy}$(-H)) field by a following description[33]

$$\Delta R_{xy} = \left( R_{xy}\left(+H\right) - R_{xy}\left(-H\right) \right) / 2 . \quad (1)$$

We find that the sign of $\Delta R_{xy}$ is unchanged regardless of $t_{NM}$ for Sample 1 (Fig. 1(b)), whereas it flips in a thick $t_{NM}$ device for Sample 2 (Fig. 1(c)).

To clarify the different change of $\Delta R_{xy}$ dependent on the sample structure, we observed $R_{xy}$ for several devices with various $t_{NM}$ in Samples 1 and 2, and References 1 and 2 at 10 K and 300 K. For a quantitative analysis of the transverse resistance, we evaluate the anomalous Hall angle. In principle, the anomalous Hall angle is defined by $\theta_{AH_0} \equiv \rho_{xy}/\rho_{xx} = \Delta R_{xy} t_{NM}/\rho_{xx}$ where $\rho_{xx}$ and $\rho_{xy}$ are the longitudinal and transverse resistivity, respectively. Due to the NM layer, however, $\theta_{AH_0}$ contains additional factors originated from the following two effects. Firstly, the current shunting flowing in the NM layer is involved in $\theta_{AH_0}$. Also a diffusing of the accumulated charge at the transverse edge to the NM layer, so-called short circuit effect[34], should be considered, The pure anomalous Hall angle $\theta_{AH}$ considering the mentioned effects reads as

$$\theta_{AH} = \frac{\Delta R_{xy} t_{NM}}{\rho_{xx}} \left( \frac{\rho_{Py}}{\rho_{NM}} \frac{t_{NM}}{t_{Py}} + 1 \right)^2 , \tag{2}$$

where $t_{Py}$ is the Py layer thickness, and $\rho_{Py}$ and $\rho_{NM}$ are the longitudinal resistivity of the Py and the NM layer, respectively. Evaluated $\theta_{AH}$ for all Samples and References at 10 K and 300 K are displayed in Fig. 2. Remarkably, $\theta_{AH}$ for References are mostly insensitive to $t_{NM}$ (Figs. 2(c), (d)). On the other hand, $\theta_{AH}$ for Samples initially changes with increasing $t_{NM}$, but its increase eventually becomes slower if $t_{NM}$ becomes sufficiently large (Figs. 2(a), (b)). Additionally, Samples 1 and 2 show different $\theta_{AH}$ dependence on $t_{NM}$ ; $\theta_{AH}$ for Sample 1 becomes larger in the thick Cu thickness regime (Fig. 2(a)), whereas that for Sample 2 becomes smaller in the thick Ag thickness regime. At the end, the sign of $\theta_{AH}$ reverses when the Ag thickness becomes sufficiently large in Sample 2 (Fig. 1(c) and Fig. 2(b)). We suppose the considerable $t_{NM}$ dependence of $\theta_{AH}$ for the Samples asserts a remarkable role of the NM/$Bi_2O_3$ interface owing to their opposite spin chirality.

For a deeper understanding of the behavior shown in Figs. 2(a) and (b), we evaluate the modulation of the pure anomalous Hall angle from the NM/$Bi_2O_3$ interface ($\theta_{AH}^{add}$) by subtracting an effect from the bulk NM and ferromagnetic metal (FM). Namely, $\theta_{AH}^{add}$ is expressed by $\theta_{AH}^{add} = \theta_{AH}$(Sample $i$)-$\theta_{AH}$(Reference $i$), $i = 1$ or $2$, where $\theta_{AH}$ (Sample $i$) and $\theta_{AH}$ (Reference $i$) corresponds to $\theta_{AH}$ for an $i$-th Sample and Reference, respectively. Since $\theta_{AH}$ for Reference is almost insensitive to the NM thickness, averaged value is used for $\theta_{AH}$ (Reference $i$). Figure 3 displays $\theta_{AH}^{add}$ for Sample 1 and 2 measured at 10 K and 300 K (Fig. 3). From the contrast trend showing in Fig. 3, we eecognize that the Cu/$Bi_2O_3$ and the Ag/$Bi_2O_3$ interfaces give an opposite influence to the transverse resistance; the Cu/$Bi_2O_3$ interface gives an increase of $\theta_{AH}$, whereas the Ag/$Bi_2O_3$ leads to the decrease of $\theta_{AH}$ with increasing $t_{NM}$. As $t_{NM}$ becomes sufficiently large, the change of $\theta_{AH}$ becomes slower. All of the mentioned trends are shown independent of temperature.

We find that the behavior of $\theta_{AH}^{add}$ shown in Fig. 3 is quite similar to the reported spin-to-charge conversion for the Cu/$Bi_2O_3$ and the Ag/$Bi_2O_3$ interfaces. As mentioned above, the Cu/$Bi_2O_3$ interface provides an opposite sign of the conversion efficiency to the Ag/$Bi_2O_3$ interface[30]. Also we find the NM thickness dependence of $\theta_{AH}^{add}$ shown in Fig. 3 is similar to the spin-to-charge conversion at the Cu/$Bi_2O_3$ interface[21, 35]. Note that the mentioned NM thickness dependences seem to originate from the enhanced interface formation with increasing of $t_{NM}$. Thus we seriously consider the spin/charge interconversion, especially the charge-to-spin conversion, at the interface makes a crucial influence on $\theta_{AH}^{add}$.

Although the transverse resistance correlates to the charge-to-spin conversion from the interface, the mechanism seems to be unclear since the generated spin current cannot make any direct influence on the transverse resistance. Also the SMR for the transverse resistance cannot be an origin of the correlation since the SMR is quadratic and insensitive to the sign of the conversion efficiency[19]. The interfacial spin Hall effect could generate a transverse spin flow with the out-of-plane spins, but it is also insensitive to the sign of the Rashba parameter[28, 36, 37].

We think the spin swapping effect in the NM layer offers a proper solution for this anomalous behavior. The spin swapping effect is belonging to the same family with the spin Hall effect (SHE). The extrinsic SHE is a spin-dependent deflection due to a scattering[38], under the relativistic magnetic field induced by the spin-orbit interaction (SOI). So far, we have focused only the spin-dependent deflection, but there is an additional effect, indeed. The scattering also leads to a spin direction alteration with a correlation with the relativistic field. Therefore, we could expect that the both the flow and the spin direction alters after the scattering. It is called spin swapping effect, and it makes a swapping between polarization and flow direction of spin current[29]. If we designate the initial spin current as $q_{ij}$, where i and j indicate the flowing and spin direction of the spin current, respectively, the spin swapping effect makes an alteration $q_{ij} \rightarrow q_{ji}$.

Owing to the opposite spin direction of the spin current from the Cu/$Bi_2O_3$ and the Ag/$Bi_2O_3$ interfaces, converted spin currents from the interfaces in Sample 1 and 2 are expressed to $q_{zy}$ and $q_{z-y}$, respectively (Fig. 4(a)). If the spin swapping effect turns on in the NM layer, the spin currents are swapped to $q_{yz}$ and $q_{-yz}$, respectively. In other words, the spin swapping effect produces a transverse spin current polarized along to the z-axis in both Samples. Here the flowing direction of the swapped spin current is opposite in Sample 1 and 2, since the initial spin direction of the spin current before the spin swapping is opposite. We find the mean free paths of Cu and Ag depends on their thickness following with a trend of the resistivity, but they are comparable or slightly shorter than NM thickness (e.g. ~9 nm for NM with $t_{NM}$=11 nm). It addresses that the spin swapping would mostly occur vicinity of the FM layer. Also we do not exclude the intermixing between the FM and the NM layers promote the spin swapping effect.

The swapped spin current may interact with the magnetization orientation of the FM layer. Similar to the giant magnetoresistance[39] and the unidirectional spin Hall magnetoresistance[23], the interaction with the FM magnetization orientation offers an influence for the spin transport. Likely, the swapped spin current feels less (more) resistive when the magnetization orientation and the spin direction of the swapped spin current is parallel (anti-parallel) (Fig. 4(b)). Under the parallel configuration, therefore, we expect more transverse spin flow along to +y (–y) direction for the Sample 1 (2). Note that the spin accumulation on the transverse side generates the charge accumulation similar to the conventional AHE since the involved electrons are all spin-polarized. As shown in above, the spin swapping effect provides a adequate mechanism for the distinctive $R_{xy}$ dependent on the spin chirality of the Rashba interface (Fig. 1(b) and (c)).

In this Letter, we have studied the transverse resistance for the Py/Cu/$Bi_2O_3$ and Py/Ag/$Bi_2O_3$ magnetic heterostructures. We found the opposite modulation of the transverse resistance dependent on the interface corresponds to the opposite spin direction of the converted spin current. We expect that the spin direction of the spin current from the interface swaps inside the NM layer, leading to the opposite modulation of the transverse resistance dependent on the spin direction. Our result also implies that the spin swapping generates a considerable influence on spin transport. Moreover, the transverse resistance provides a good platform to observe the spin swapping.

This work is supported by a Grant-in-Aid for Scientific Research on the Innovative Area, “Nano Spin Conversion Science” (Grant No. 26103002), and Japan Society for the Promotion of Science through Program for Leading Graduate Schools (MERIT)..

yotani@issp.u-tokyo.ac.jp; junyeon.kim@riken.jp

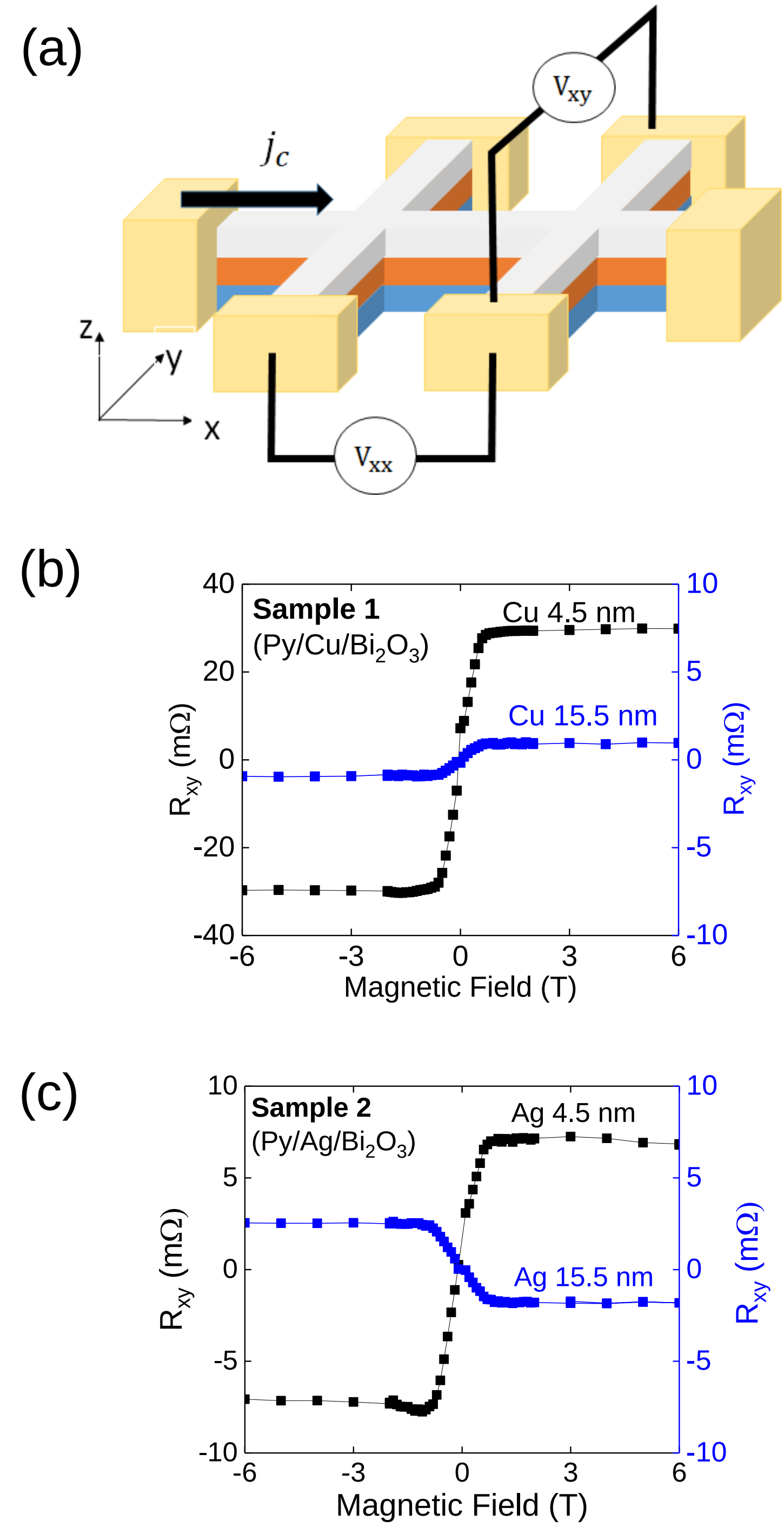

**Fig. 1.** (a) Scheme of the measurement. (b),(c) Typical $R_{xy}$ for Sample 1 (b) and Sample 2 (c) at room temperature. The magnitude of $R_{xy}$ for a thin Cu device (Black square) is displayed in the left axis, whereas that for a thick Cu device (Blue square) corresponds to the right axis. All data were obtained under room temperature.

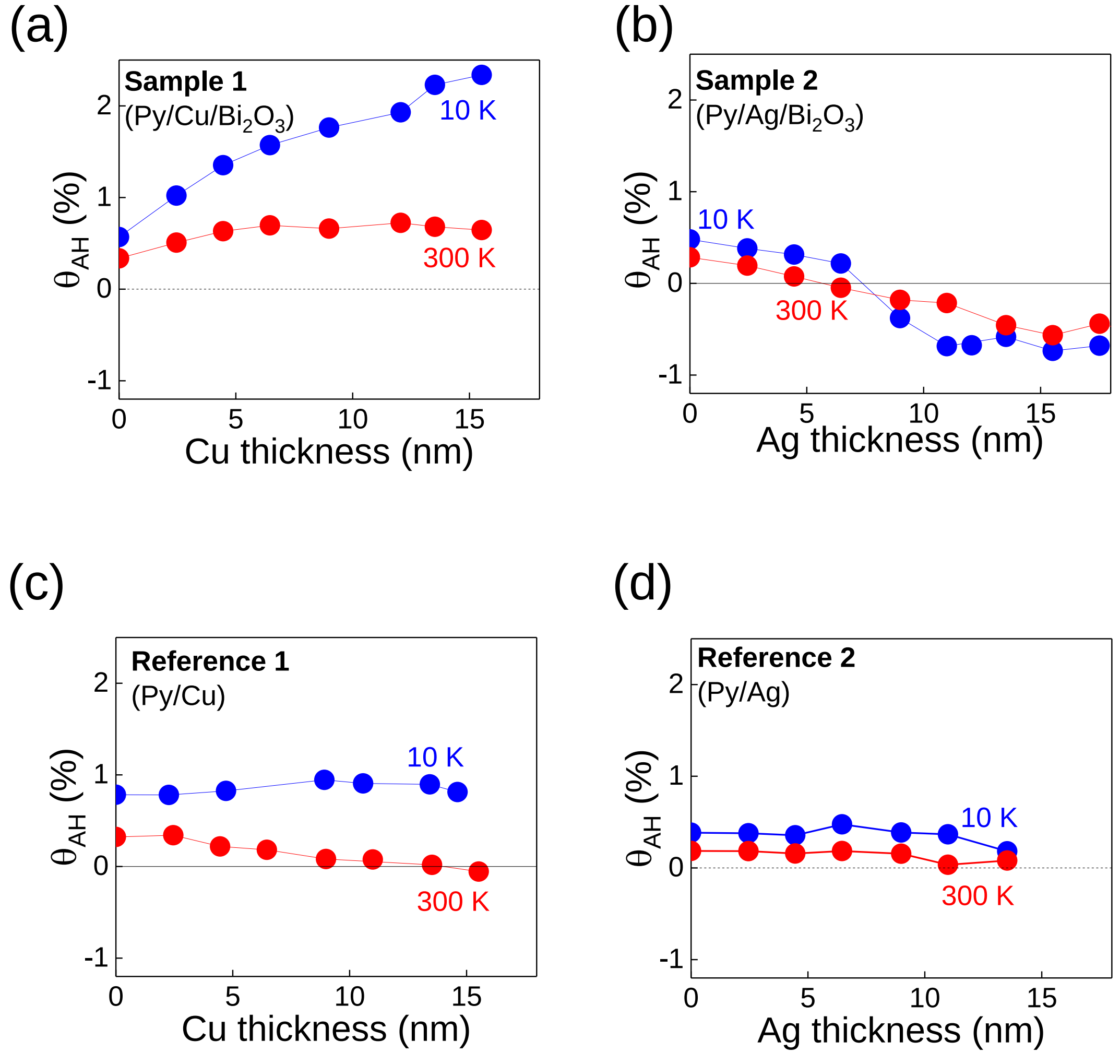

**Fig. 2.** (a),(b) $\theta_{AH}$ as a function of $t_{NM}$ at 10 K (blue circle) and 300 K (red circle) for Sample 1 (a) and Sample 2 (b). (c),(d) $\theta_{AH}$ as a function of $t_{NM}$ at 10 K (blue circle) and 300 K (red circle) for Reference 1 (c) and Reference 2 (d).

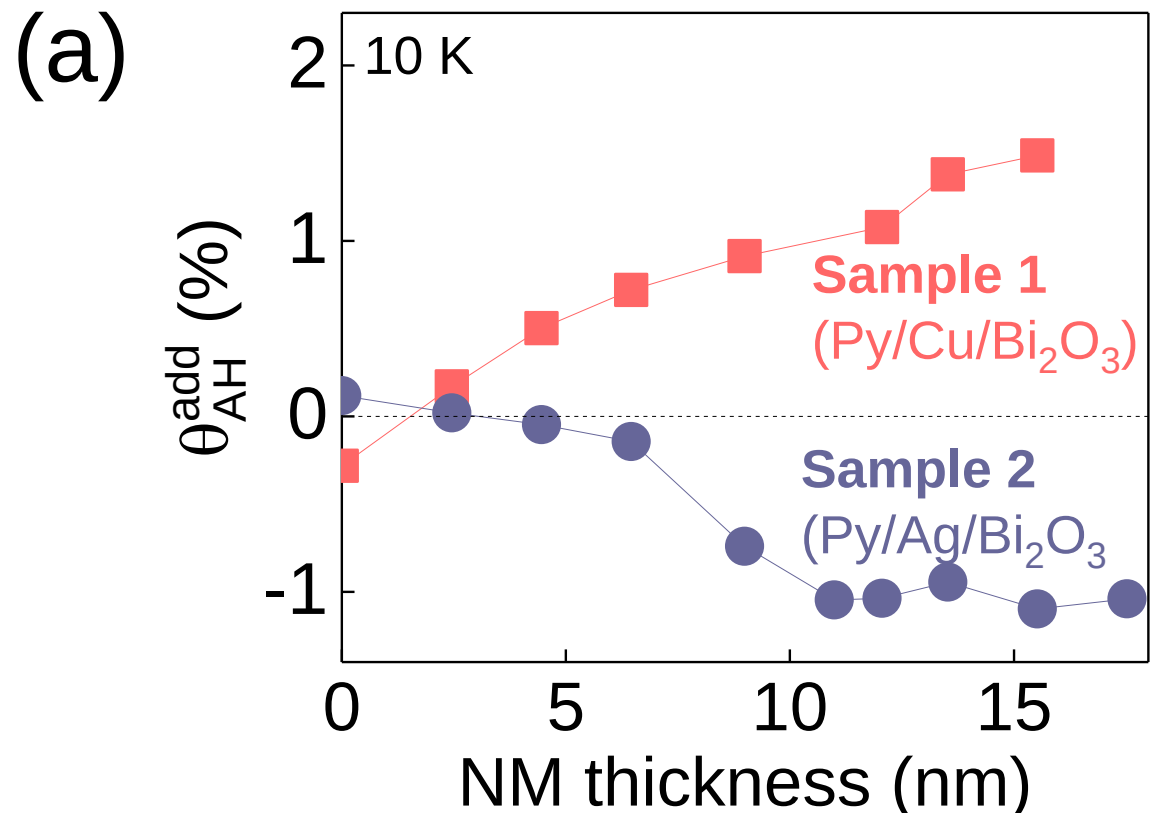

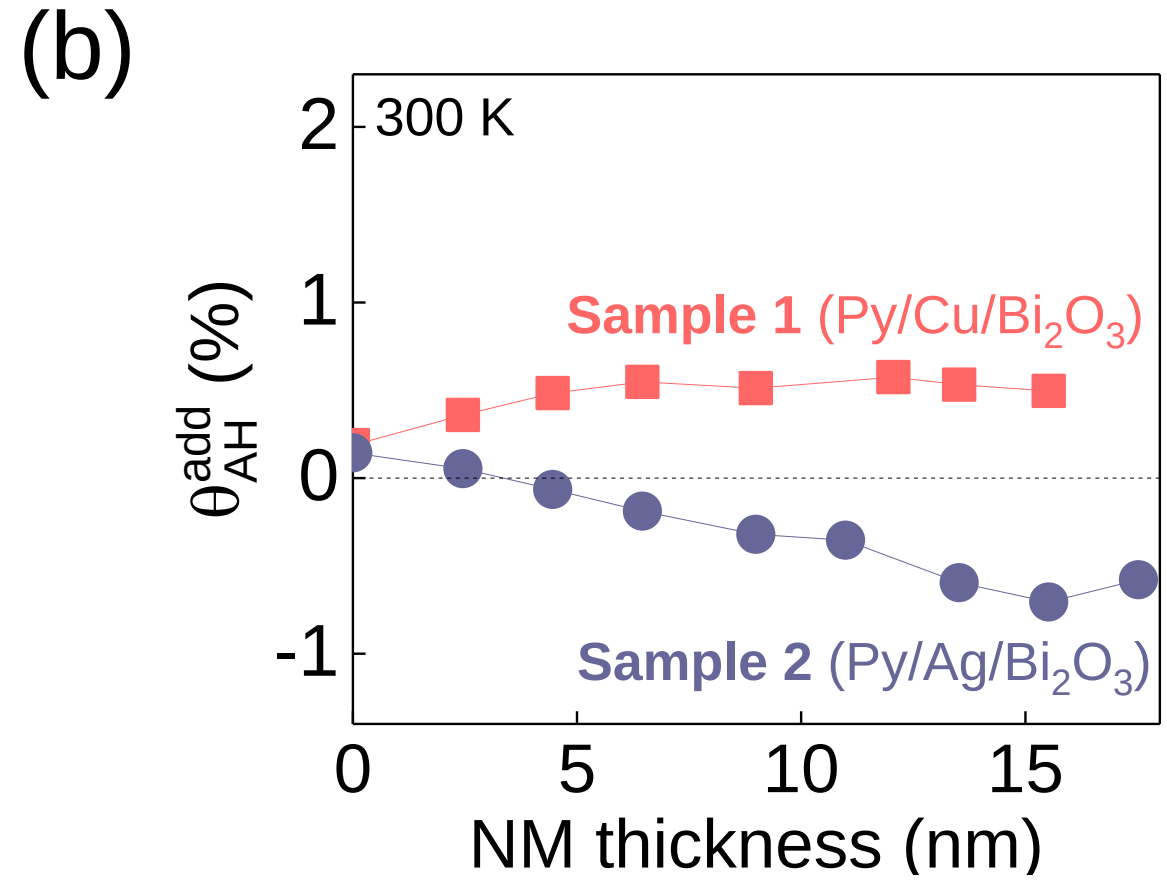

**Fig. 3.** (a),(b) $\theta_{AH}^{add}$ for Sample 1 (pink square) and Sample 2 (gray circle) at 10 K (a) and 300 K (b).

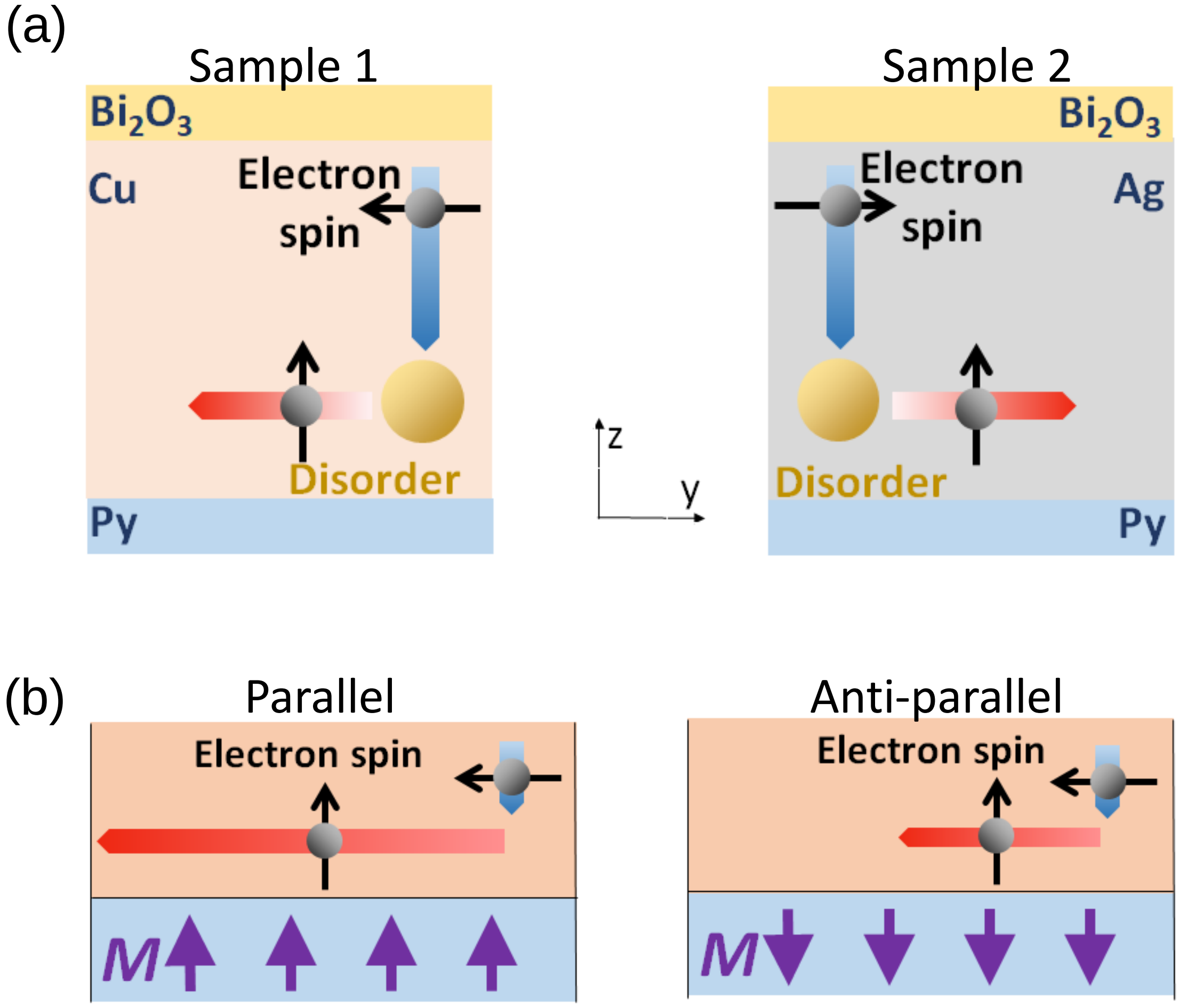

**Fig. 4.** (a) Scheme of the spin swapping in Cu (left panel) and Ag (right) layer. (b) FM magnetization orientation dependent spin transport after the spin swapping: Spin direction of the spin current is parallel (left panel) and anti-parallel (right panel) to the FM magnetization orientation.